\documentclass[12pt,showpacs]{revtex4}

\usepackage[english]{babel}
\usepackage{amsbsy}
\usepackage{amstext}
\usepackage{graphicx}
\usepackage{amssymb}

\newcommand {\be} {\begin{eqnarray}}
\newcommand {\ee} {\end{eqnarray}}
\newcommand {\ba} {\begin{array}}
\newcommand {\ea} {\end{array}}
\newcommand {\bdm} {\begin{displaymath}}
\newcommand {\edm} {\end{displaymath}}

\begin{document}

\title{Wavelet entropy filter and cross-correlation of gravitational wave data}
\author{R~Terenzi$^{1,2}$ and R~Sturani$^3$}
\affiliation{$^1$ INFN, Sezione di Roma Tor Vergata, I-00133 Roma,~Italy}
\affiliation{$^2$ Istituto di Fisica dello Spazio Interplanetario (IFSI) INAF, 
I-00133 Roma,~Italy}
\affiliation{$^3$ Istituto di Fisica, Universit\`a di Urbino, I-61029 Urbino,
~Italy}
\affiliation{INFN, Sezione Firenze/Urbino, I-50019 Sesto Fiorentino,~Italy}
\email{roberto.terenzi@cern.ch,riccardo.sturani@uniurb.it}

\begin{abstract}
We present a method for enhancing the cross-correlation of gravitational wave 
signals eventually present in data streams containing otherwise uncorrelated
noise. Such method makes use of the wavelet decomposition to cast 
the cross-correlation time series in time-frequency space. Then an entropy 
criterion is applied to identify the best time frequency resolution, i.e.
the resolution allowing to concentrate the signal in the smallest number of
wavelet coefficients. By keeping only the coefficients above a certain 
threshold, it is possible to reconstruct a cross-correlation time series 
where the effect of common signal is stronger.
We tested our method against signals injected over two data streams of 
uncorrelated white noise. 
\end{abstract}


\maketitle

\section{Introduction}
\label{se:intro}

Gravitational waves (GW) are expected to be emitted by a variety of 
cosmological and astrophysical sources and several detectors are now 
operating to directly observe such a signal. 

Some sources has been thoroughly studied and the expected GW-signal they emit 
is well modeled, like binary coalescences (without spin) and spinning neutron 
stars. For this class of sources the standard, matched-filtering techniques 
can be adopted, which are able to dig well under the noise floor. 
However some astrophysical systems may emit signals which are poorly modeled 
at the moment, not to mention unknown sources. This is the case for spinning 
coalescing systems, stellar core collapse events, cosmological stochastic 
signal, to mention just a few, known cases.

In ordered to be able to spot these unmodeled signals it is necessary to use a 
wide net, capable of detecting any suspect excess noise in the data streams.
Following this strategy several burst searches are performed by the LIGO/VIRGO
\cite{ligovirgo} as well as by the resonant antennas network 
\cite{Vinante:2006uk,Astone:2008zz}, in which the 
technique of \emph{cross-correlation} of data streams  between two (or more) 
detectors is employed, possibly jointly with other techniques, to identify 
GW-event candidate, see e.g.\cite{laura,Abbott:2007wu,Acernese:2007ek}. 
The cross-correlation analysis consists in obtaining a suitable 
defined data stream out of two detector data streams, and check for anomalous
excesses in it. Cross-correlation can also be used in combination with 
matched filtering technique.

Here we present a method to enhance to capability of cross-correlating
techniques to detect GW-signals. Our method is based on a \emph{wavelet} 
decomposition of the cross-correlation data stream. Wavelet decomposition
have been already adopted in GW-data analysis, see e.g. \cite{Moh,Syl}. 
In particular the \emph{WaveBurst} algorithm \cite{Klim,Klim2}
has been employed in the the LIGO/VIRGO burst search. 

Wavelet decomposition is a time-frequency decomposition. Unlike in the familiar
case of Fourier Transform, in wavelet decomposition the time and 
frequency resolution is a free parameter (though the time and frequency 
resolutions are not independent on each other). The method described in the 
following section is able to \emph{adaptively} select the decomposition level, 
or time-frequency resolution, which is best for the generic data stream 
analyzed.
In particular this method, presented in a different context in 
\cite{Sturani:2007tc}, picks the resolution which gathers the power of the 
data stream in the smallest possible number of wavelet 
coefficients, enhancing the possibility of singling out excesses of 
cross-correlation. The resolution choice is obtained by considering an 
\emph{entropy} function \cite{Wic}, which is maximum when the power of the 
signal is equally distributed among all the coefficients and minimum when all 
the power is concentrated in only one coefficient.
Once the best resolution has been individuated, by a process of thresholding 
the detection of correlation excesses can be enhanced with respect to the 
analysis obtained by simply cross-correlating the data streams without further 
processing. 

The paper is organized as follows: in sec.~\ref{se:method} we describe the 
method, in sec.~\ref{se:result} we report the results obtained by analyzing 
a stream of white data sampled at 5kHz containing injections of short-duration 
signals. 
Sec.~\ref{se:concl} contains the conclusion of our analysis.

\section{The method: wavelet packet and entropy filter}
\label{se:method}

A wavelet transform is used to transform a time series into a mixed 
time-frequency one. 
Given an initial series made of $N$ data points, with sampling $\Delta t_0$, 
it contains information on frequencies ranging from zero up to the Nyquist 
frequency $f_{max}=(2\Delta t_0)^{-1}$.
Different decomposition \emph{levels} are available in a wavelet transform, 
labeled 
by a (limited) integer number $j$. At each level the wavelet-transformed 
data stream contains $N$ coefficients as the initial time series, arranged in 
$2^j$ \emph{layers}, each with $N\times 2^{-j}$ coefficients. At level $j$ the 
time resolution is $\Delta t_j=2^{j}\Delta t_0$ and the frequency resolution 
is $\Delta f_j=2^{-j}f_{max}=1/(2\Delta t_j)$. The coefficients in the 
$i^{\rm th}$ layer ($0<i<2^j$) refer to the frequency bins characterized by 
$i2^{-j}\leq f/f_{max}<(i+1)2^{-j}$.

\begin{figure}[th]
\begin{center}
\includegraphics[width=.8\linewidth]{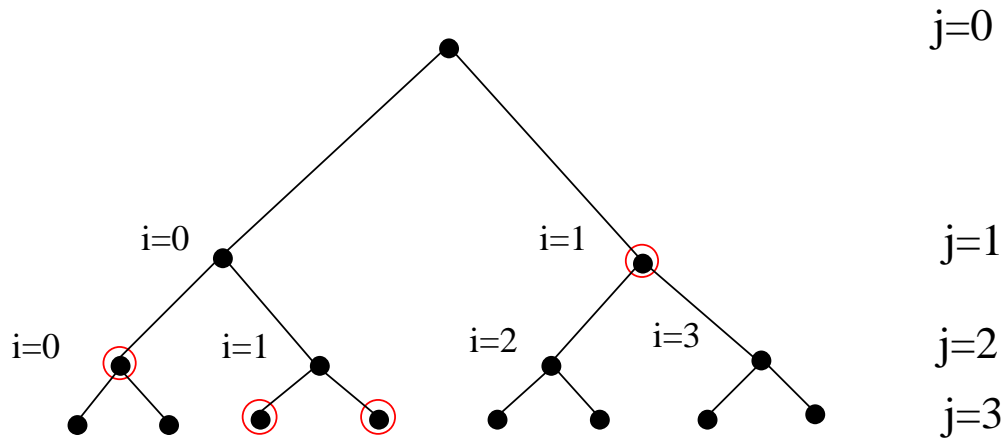}
\caption{Schematic example of a binary tree for a wavelet decomposition up
  to level 3. The layers in each level are explicitly labeled for $j<3$. 
  The red circles show an example of a possible choice for a tree. 
  In this examples it is made by choosing the subspaces with 
  $\{j,i\}=\{(2,0),(3,2),(3,3),(1,1)\}$, from \cite{Sturani:2007tc}.}
\label{fig:tree}
\end{center}
\end{figure}

The output of a wavelet transform can be arranged in a binary tree where each 
node is an orthogonal vector subspace at level $j$ and at \emph{layer} 
$i$. Together the layers of the same level contain exactly the same 
information of the original time series, so that complete reconstruction of 
the original signal is possible by collecting the coefficients of any level. 
Anyway this is not the only possible choice to reconstruct the original 
time-signal: it can be shown (see e.g. \cite{Mallat}) that other 
{\it admissible trees} $T$ completely represents 
the original signal in the wavelets domain, where an admissible tree is a 
sub-tree of the original binary decomposition tree where every node has 
exactly $0$ or $2$ children nodes, see e.g. fig.~\ref{fig:tree}, where the
nodes of an example of an admissible tree are marked by red circles. We 
denote an admissible tree by $T_{\{k,l\}}$, as it implies a choice of a set 
of pairs of level/layers. \\
Having such a redundancy, how to choose the set of  subspaces to represent 
the signal? The standard choice is to take all the layers of a single 
decomposition level, but even in this case one has the freedom to pick among 
several levels.

The way we propose to choose in the set of admissible trees $T_{\{k,l\}}$ is 
based on an {\it entropy } function $E_T$ defined as follows:
\begin {equation}
E_{T_{\{k,l\}}}= -\sum_{i,j\in \{k,l\}} {x^{2}_{ij} 
\over{||X||_T^{2}}} \log {x^{2}_{ij} \over{||X||^{2}_T}}\,,
\label{eq:entropy}
\end{equation}
where the $x_{ij}$ are the wavelet coefficients of a level $j$ and layer $i$ 
and 
\begin {equation}
||X||^{2}_T= \sum_{i,j\in \{k,l\}} {x_{ij}^{2}}\,.
\end{equation}

From the complete decomposition tree, we select among all the possible 
admissible trees the one which have the minimum cost, i.e. the $\bar T$ for 
which $E_{\bar T}$ is minimum, following the {\it entropy criterion} 
(\cite{Wic} where a different notation is used). 
From among all the admissible binary trees, $\bar T$ is the one that 
represents the signal most efficiently, as it uses the least possible number 
of wavelet coefficients.\\
For instance in the case in which only one of the $x_{ij}$'s is non-vanishing 
(maximal concentration) the entropy function is vanishing.
On the other hand, if for some tree the decomposition coefficients are all 
equal, say $x_{ij} = 1/N$, the entropy in this case is maximum, $\log N$.

After having performed the wavelet transform (a Symlet basis has been used) 
and the evaluation of the entropy function to determine the best tree, for 
each node in the resulting tree we set a threshold $\theta_{i,j}$ equal to a 
fraction of the value of the maximum wavelet coefficient $w_{i,j}$ in the node.
Then a {\em soft} 
threshold function has been applied, i.e. each coefficient in the node has been
changed according to the following function $f$:
\begin{equation}
  \begin{tabular}{ll}
    $f(w_{i,j}(k))=0$ & if $w_{i,j}(k)<\theta_{i,j}$ \\
    $f(w_{i,j}(k))=w_{i,j}(k)-\theta_{i,j}$ & if $w_{i,j}(k)\geq\theta_{i,j}$
  \end {tabular}
\end{equation}
where $k$ ($0\leq k<N2^{-j }$) runs over the coefficients of the layer.

Then the time series has been recomposed from this mutilated wavelet 
coefficient set, where spurious disturbances (non concentrated in 
time and frequency) have been epurated. This will allow, as it is shown in the 
next section, to single out more clearly the excess in correlation due to 
actual signals.
The wavelet decomposition has then been used to apply a non-linear filter
to the r-series.

\section{Results}
\label{se:result}
The cross correlation $r$ of two time series $\{x_i\},\{y_i\}$, each of
length $N$, is given by
\be
\label{xcorr}
r=\frac{\sum_{i=1}^N(x_i-\bar x)(y_i-\bar y)}
{\sqrt{\sum_{i=1}^N(x_i-\bar x)^2}\sqrt{\sum_{i=1}^N(y_i-\bar y)^2}}\,,
\ee
where an overbar stands for average value.
Such quantity measures the correlation between two data streams, as it would 
be produced by a common GW signal embedded in uncorrelated detector noise 
\cite{laura} and it compares waveforms without being sensitive to their 
relative amplitude. In real cases, data are usually filtered and/or 
whitened before applying the cross-correlation analysis. We have simulated
two series of uncorrelated, white noise data.
To test if our method is able to enhance the cross correlation output we 
have injected common signals of different amplitude and shape into the two
data streams.
For simplicity we have concentrated our attention on one kind of signal, an 
exponential-sine, whose time profile is given by:
\be
\label{t_profile}
h(t)=h_0\sin(2\pi f_0 t)e^{-t/\tau}\qquad t>0\,,
\ee
(and $h(t)=0$ for $t<0$) for two different amplitudes, central frequency 
$f_0=920$Hz and $\tau={10,100}$msec.
The amplitudes have been chosen so to reproduce an $SNR=0.47,0.71$,$2.24$ and 
$1.48$, where the SNR is defined as 
\be
\label{snr}
SNR\equiv \sigma_s/\sigma_n\,,
\ee
where $\sigma_n$ is the noise standard deviation and $\sigma_s$  the signal 
standard deviations computed over a $5\tau$ period from the injection on. 
The correlation time-windows (determined by $N$ in eq.~(\ref{xcorr}) and by 
the sampling time) are taken to be respectively $100$, $500$ msec for the 
$\tau=10,100$ msec injections.

\begin{figure}
  \begin{center}
    \includegraphics[width=.8\linewidth]{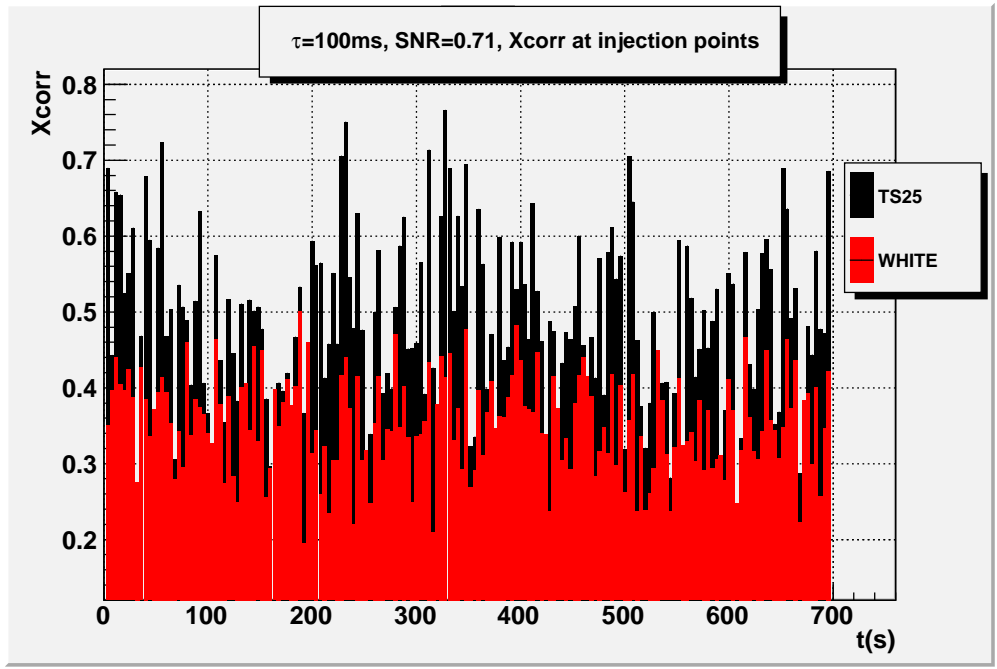}
    \caption{Example of the cross-correlation of data streams for identical 
      injections of the type (\ref{t_profile}) with $f_0=920$Hz,$\tau=100$ 
      msec, correlation time-window of $500$ msec, $SNR=0.71$, as defined in 
      eq.~(\ref{snr}). 
      The figure displays the maximum value of the $r$-coefficient for each 
      injection (separated by 4 secs), before (red) and after (black) the 
      wavelet entropy filter has been applied.}
    \label{example}
  \end{center}
\end{figure}

As an example, we report in fig.~\ref{example} the values of the maximum of
the $r$-coefficients in correspondence of a series of injections 
(\ref{t_profile}).

\begin{figure}
  \begin{center}
    \includegraphics[width=.8\linewidth]{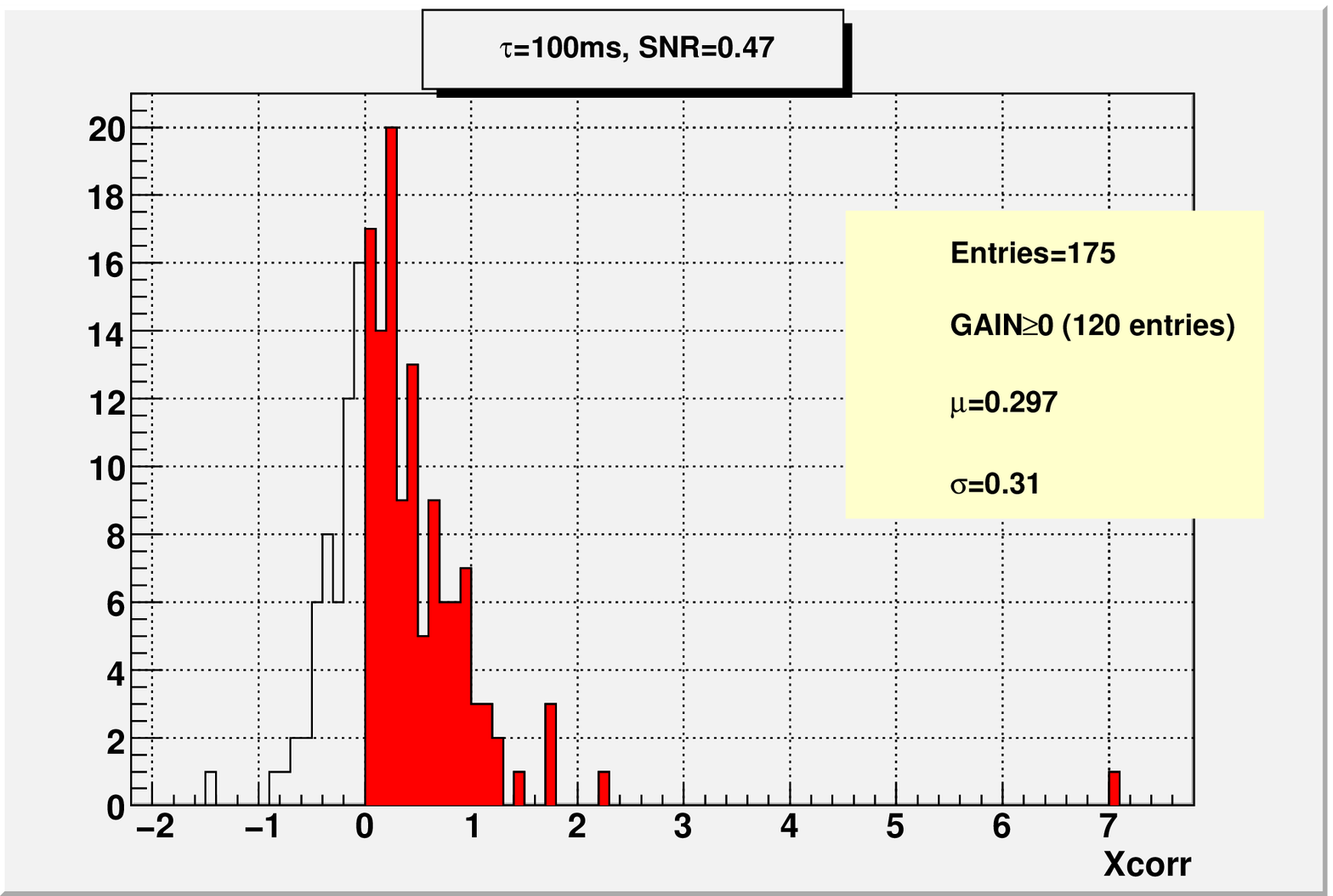}
    \caption{Histogram of gain factor, as defined by eq.~(\ref{gain}) and 
      correspondent average $\mu$ and standard deviation $\sigma$. 
      Parameters $SNR$ and $\tau$ of the injections
      are displayed, $f_0=920$Hz.}
    \label{histo1}
  \end{center}
  \begin{center}
    \includegraphics[width=.8\linewidth]{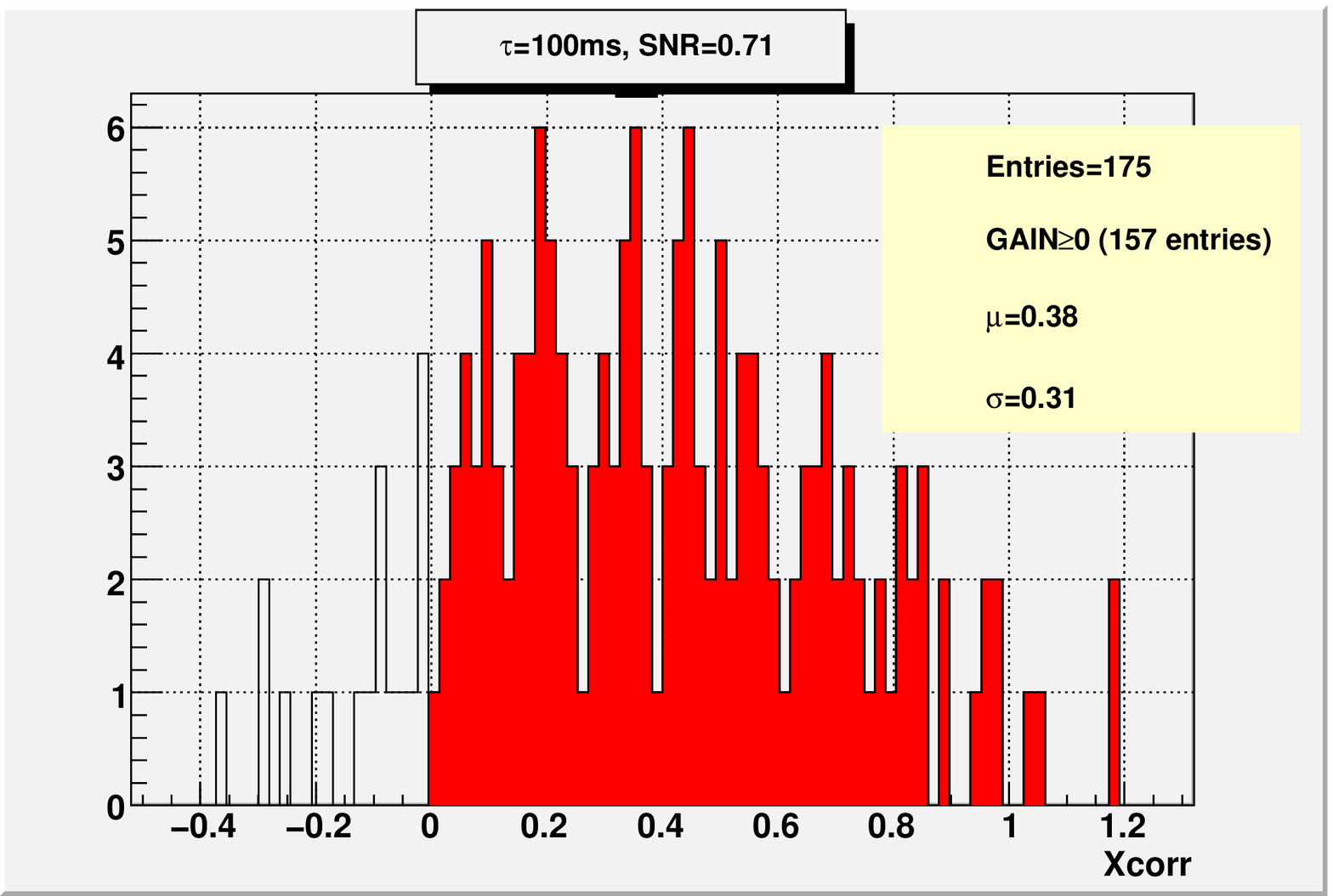}
    \caption{Histogram of gain factor. Parameters $SNR$ and $\tau$ are
      displayed, $f_0=920$Hz.}
    \label{histo2}
  \end{center}
\end{figure}

\begin{figure}
  \begin{center}
    \includegraphics[width=.8\linewidth]{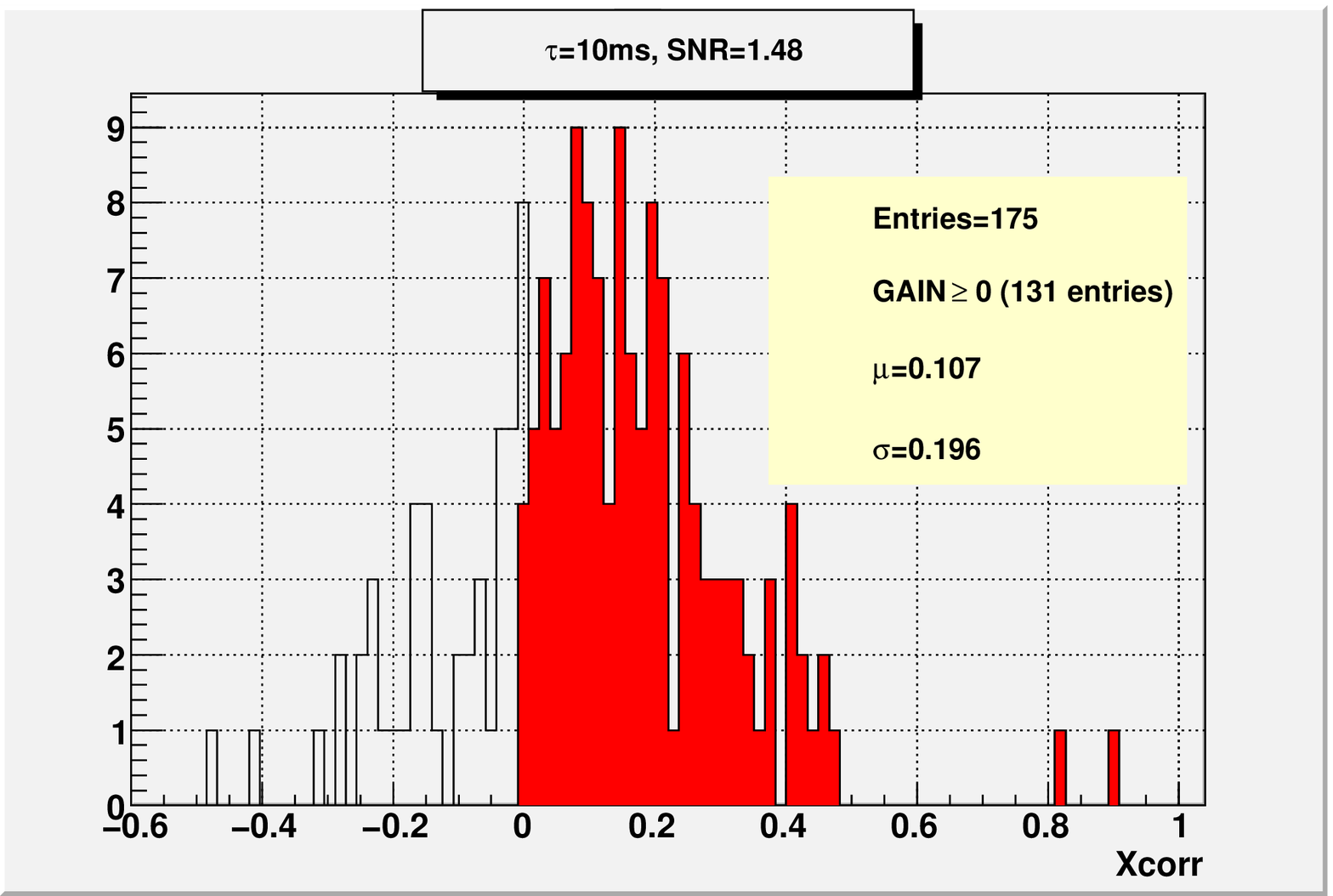}
    \caption{Histogram of gain factor. Parameters $SNR$ and $\tau$ are
      displayed, $f_0=920$Hz.}
    \label{histo3}
  \end{center}
  \begin{center}
    \includegraphics[width=.8\linewidth]{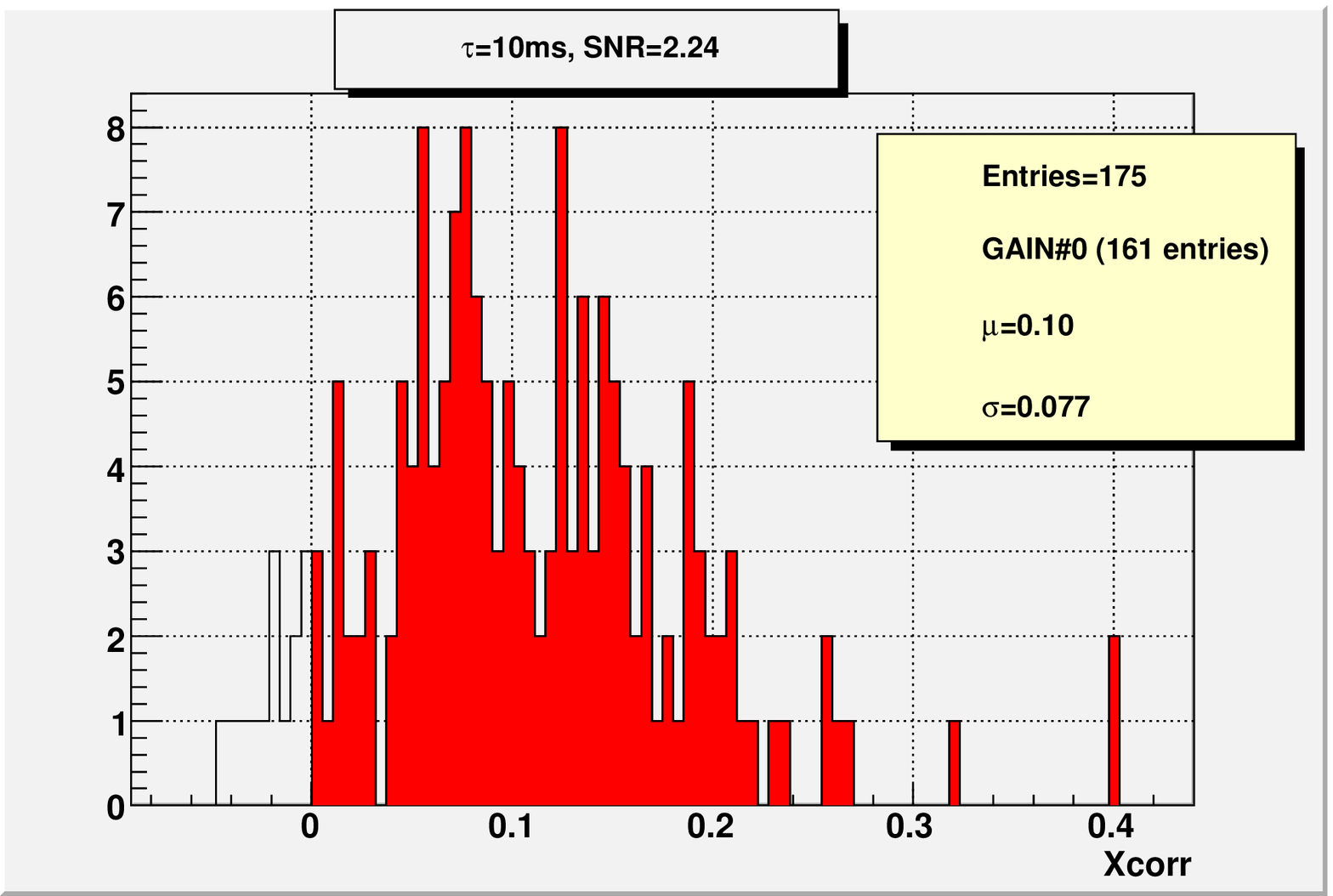}
    \caption{Histogram of gain factor. Parameters $SNR$ and $\tau$ are
      displayed, $f_0=920$Hz.}
    \label{histo4}
  \end{center}
\end{figure}

In order to quantify the enhancement due to the entropy filter we define a 
\emph{gain} factor by considering, for each injection, the maximum 
value\footnote{The maximum is computed over a 200msec time interval
centered at the injection time.} of the $r$-coefficient for the data before 
($\bar r$) and after ($\bar r_W$) the treatment with the wavelet 
entropy filter and then define the quantity
\be
\label{gain}
{\rm gain}=\frac{\bar r_W-\bar r}{\bar r}\,.
\ee
Figs.~\ref{histo1}-\ref{histo4} show that for different choice of the injection
parameters the wavelet entropy filter thus indeed lead to a statistical 
enhancement of the cross-correlation, the enhancement being stronger for 
stronger signals. 

In order to show that this method let excesses of correlation step out of the 
noise floor more clearly, while not introducing any significant increase of 
false alarm rate, we have evaluated the average and standard deviation 
of the maximum of the $r$-coefficients obtained by cross-correlating
white data without any injection before and after the application of the 
entropy filter. The result summarized in tab.~\ref{statistica} shows that in 
absence of signal the application of the wavelet entropy filter is 
irrelevant.\\
On the other hand, analyzing the cross-correlation time series obtained by 
data \emph{with} injections, and counting how many 
$\bar r,\bar r_W$-coefficients have exceeded the average by more than 
3$\sigma$ and 4$\sigma$ ($\mu$ and $\sigma$ reported in 
tab.~\ref{statistica}), we have obtained the results reported in 
tab.~\ref{excess}, showing that the wavelet entropy filter 
is powerful in enhancing the detection of signals embedded into noise.

\begin{table}[t]
  \begin{center}
    \begin{tabular}{|c|c|c|c|c|}
      \hline
      Correlation time-window & \multicolumn{2}{|c|}{$\bar r$}
      & \multicolumn{2}{|c|}{$\bar r_W$}\\
      \cline{2-5}
      (msec) &  $\mu$ & $\sigma$ & $\mu$ & $\sigma$ \\
      \hline
      $100$ & 0.21 & 0.14 & 0.21 & 0.15 \\
      \hline
      $500$ & 0.041 & 0.081 & 0.040 & 0.075 \\
      \hline
    \end{tabular}
    \caption{Background mean value $\mu$ and standard deviation $\sigma$
      of the maximum cross-correlation coefficients obtained from white data 
      (no injections) before ($\bar r$) and after ($\bar r_W$) the application 
      of the entropy filter. The maximum is computed over a time-interval of 
      $200$ msec.}
    \label{statistica}
  \end{center}
  \begin{center}
    \begin{tabular}{|c|c|c|c|c|c|}
      \hline
      \multicolumn{2}{|c|}{} & \multicolumn{2}{|c|}{$\tau=10$ msec}
      & \multicolumn{2}{|c|}{$\tau=100$ msec}\\
      \cline{3-6}
      \multicolumn{2}{|c|}{} & SNR=2.24 & SNR=1.48 & SNR=0.71 & SNR =0.48 \\
      \hline
      $4\sigma$ & \#$\bar r_W$ & 127 & 14 & 153 & 38 \\
      \cline{2-6}
      & \#$\bar r$ & 84 & 5 & 84 & 2 \\
      \hline
      $3\sigma$ & \#$\bar r_W$ & 172 & 91 & 172 & 73 \\
      \cline{2-6}
      & \#$\bar r$ & 169 & 70 & 151 & 28 \\
      \hline
    \end{tabular}
    \caption{Number (out of 175 injections) of $\bar r$ and 
      $\bar r_W$-coefficients at injection time exceeding by $3\sigma$ and 
      $4\sigma$ the mean value of the $r$-coefficients of data without 
      injections. The time-window for correlation are taken $100,500$ ms for 
      $\tau=10,100$ ms respectively.}
    \label{excess}
  \end{center}
\end{table}

\section{Conclusions}
\label{se:concl}
We have presented a method that makes use of a wavelet decomposition to 
determine which part of the cross-correlation time series can be zeroed 
without loosing interesting feature of the signals, thus reducing the impact 
of noise and we showed that this method can enhance the detection effciciency
of signal embedded into noise while not increasing the false alarm rate.
We plan to release the code to the gravitational wave community in a near 
future.

\section*{Acknowledgements}
The authors wish to thank the ROG collaboration for useful discussions.

\end{document}